\documentclass{article}
\usepackage{ijcai26}

\usepackage{times}
\usepackage{soul}
\usepackage{url}
\usepackage[hidelinks]{hyperref}
\usepackage[utf8]{inputenc}
\usepackage[small]{caption}
\usepackage{graphicx}
\usepackage{amsmath}
\usepackage{amsthm}
\usepackage{amssymb}
\usepackage{booktabs}
\usepackage{algorithm}
\usepackage{algorithmic}
\usepackage[switch]{lineno}
\usepackage{xcolor}
\usepackage{hyperref}

\title{Kaiwu-PyTorch-Plugin: Bridging Deep Learning and Photonic Quantum Computing for Energy-Based Models and Active Sample Selection}
\author{
    Hongdong Zhu$^{1,^\dagger}$\and
    Qi Gao$^{1,^\dagger}$\and
    Yin Ma$^{1,^\dagger}$\and
    Shaobo Chen$^1$\and
    Haixu Liu$^1$\and
    Fengao Wang$^2$\and
    Tinglan Wang$^1$\and
    Chang Wu$^1$\And
    Kai Wen$^{1,*}$
    \affiliations
    $^1$Beijing QBoson Quantum Technology Co., Ltd.
    $^2$Chinese Academy of Sciences\\
    \emails
    \{zhuhd, gaoq, may, chensb, liuhx, wangtl, wuc, wenk\footnote{Corresponding Author. 
$^\dagger$These authors contributed equally.
}\}@boseq.com,
    wangfengao21@mails.ucas.ac.cn
}

\begin{document}

\maketitle

\begin{abstract}
    This paper introduces the Kaiwu-PyTorch-Plugin (KPP) to bridge Deep Learning and Photonic Quantum Computing across multiple dimensions. KPP integrates the Coherent Ising Machine into the PyTorch ecosystem, addressing classical inefficiencies in Energy-Based Models. The framework facilitates quantum integration in three key aspects: accelerating Boltzmann sampling, optimizing training data via Active Sampling, and constructing hybrid architectures like QBM-VAE and Q-Diffusion. Empirical results on single-cell and OpenWebText datasets demonstrate KPP’s ability to achieve SOTA performance, validating a comprehensive quantum-classical paradigm. 
\end{abstract}
\section{Introduction}
The success of deep learning is largely attributed to the predictability provided by Scaling Laws, as well as the high degree of alignment between the gradient descent backpropagation algorithm and large-scale parallel computing hardware (such as GPUs). However, for a class of models with powerful unsupervised learning capabilities—Energy-Based Models (EBMs), such as Restricted Boltzmann Machines (RBMs) \shortcite{ackley1985learning} and Boltzmann Machines (BMs) \shortcite{hinton2002training} —the training process on classical hardware relies heavily on Gibbs Sampling from high-dimensional energy landscapes to estimate gradients. As the number of nodes increases, Gibbs sampling becomes highly inefficient and prone to getting trapped in local minima. Moreover, with the mass expansion of datasets, the infiltration of data toxicity and redundant information significantly diminishes the information gain per sample.

In this context, Quantum Computing , particularly the Coherent Ising Machine (CIM) \shortcite{inagaki2016coherent}, has emerged as a physics-driven computational paradigm with the potential to further unlock the efficiency and performance of deep learning. Unlike universal gate-model quantum computing, CIM focuses on solving the ground state and thermal distribution sampling of the Ising Model. These physical properties share a profound mathematical isomorphism with $NP$-hard combinatorial optimization problems in deep learning, such as EBM training, active learning, and sample selection in information retrieval \shortcite{pasin2025quantumclef}. To date, the integration of CIM-based Quantum Computing with deep learning has demonstrated superiority in tasks such as single-cell representation, biomolecular sequence generation and design, and text generation.

QAML\shortcite{pinilla2024quantum} and the D-Wave-PyTorch-Plugin \shortcite{winci2020path} both leverage the Ocean library developed by D-Wave to provide annealing sampling algorithms that accelerate the training of EBMs. However, D-Wave’s superconducting quantum annealers require an operating environment near absolute zero, and their processor chips utilize a sparse topology, which limits their ability to directly map fully connected layers. Furthermore, these two frameworks focus exclusively on a narrow set of scenarios—specifically, utilizing quantum annealing sampling to speed up the training of energy models. 

To bridge the gap between quantum algorithms, specialized hardware, and deep learning architectures, we introduce Kaiwu-Pytorch-Plugin \footnote{The video is available at \href{https://drive.google.com/file/d/1YcbooLvwOc2qf5vqrn_X5f7uAmmzBlLZ/view?usp=sharing}{\textcolor{magenta}{Google Drive}}}. As a PyTorch extension tailored for CIM, it offers efficient, stable, and secure cloud-native quantum acceleration. The plugin integrates Boltzmann Sampling to significantly accelerate energy model convergence and introduces an Active Sampling strategy powered by global batch-level optimization. 
\section{System Architecture}
\subsection{Handware}
The CIM is an innovative computational architecture that leverages the natural evolution of optical physical systems to solve complex combinatorial optimization problems. It consists of three core modules: optical quantum state preparation, optical quantum memory, and an integrated measurement and control system \shortcite{wei2026versatile}.
Specifically, the system utilizes a sequence of laser pulses circulating within a 1km fiber loop as the computational carrier, where the phase of each pulse ($0$ or $\pi$) represents a binary variable in the optimization problem \shortcite{wang2013coherent}. Under the regulation of a measurement-feedback mechanism, the system reads the pulse states via photodetectors and transmits them to an FPGA. The FPGA calculates the mutual interactions between variables in real-time based on the mathematical model of the target problem (i.e., the Ising model matrix). These interactions are then converted into feedback optical signals and reinjected into the optical path to interfere with the original pulses. 
As the system's underlying energy (pump light) gradually increases, the initially random network of optical pulses---under the continuous constraint of feedback signals---spontaneously evolves toward a state of minimal global conflict and lowest energy (a physical process known as spontaneous symmetry breaking). When the system reaches stable oscillation, the resulting phase configuration of the pulse population represents a high-quality or optimal solution to the original combinatorial optimization problem. This architecture ingeniously transforms a complex mathematical search process into the natural convergence of a physical system, achieving microsecond-level acceleration for NP-hard problems. Compared to superconducting quantum annealers \shortcite{kadowaki1998quantum}, the CIM can operate stably at room temperature with lower costs and more easily achieves all-to-all connectivity between variables \shortcite{mcmahon2016fully}. Furthermore, compared to technical routes such as trapped ions \shortcite{haffner2008quantum}, CIMs supports orders of magnitude more qubits.
\begin{figure}[htbp]
    \centering
    \vspace{-1em} 
    \includegraphics[width=0.48\textwidth]{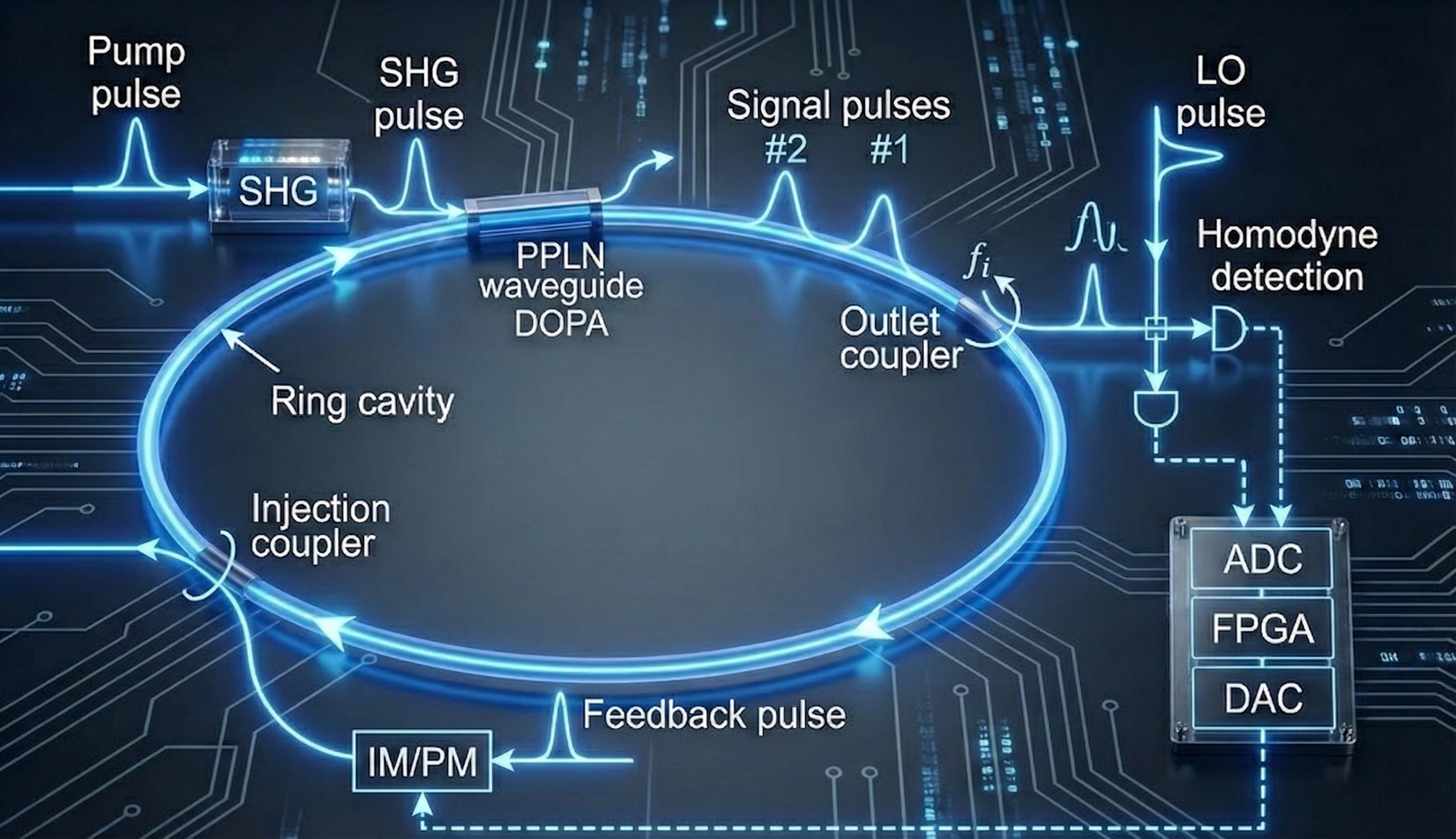}
    \vspace{-1.5em} 
    \caption{Compact view of the CIM structure.}
    \label{fig:cim_structure}
    \vspace{-1.5em} 
\end{figure}
\vspace{-0.5em}
\subsection{Datasets} 
To demonstrate the versatility and superiority of our method, we integrate two main sets of datasets from vastly different domains into our framework.

\paragraph{NLP.} OpenWebText \shortcite{radford2019language}, known for its high linguistic diversity and broad topical coverage, is a widely used foundation for training general LLMs. Here, it is utilized to evaluate the model's generative capabilities in natural language.

\paragraph{Bio.} We select two small-scale single-cell datasets from scvi-tools (PBMC and Pancreas) \shortcite{ergen2025scvi}, alongside the medium-scale Human Fetal Lung Cell Atlas \shortcite{he2022human} and the large-scale HLCA Core \shortcite{sikkema2023integrated}. These are used to assess the generalization and robustness of our method across datasets of varying scales and functions.

We also integrate the MNIST digit dataset and a human RNA dataset, providing corresponding models and training examples for developer exploration. 

\subsection{Dataloader and Active Selection} 
To ensure efficient sample management, we construct a vector database to store high-dimensional representations of all training data. Selective retrieval of sample embeddings is implemented by overriding the \texttt{\_\_getitem\_\_} method of the PyTorch \texttt{DataLoader}. Furthermore, we design a custom \texttt{collate\_fn} to model sample screening as an Active Sample Selection problem, which is formulated as minimizing the following Quadratic Unconstrained Binary Optimization (QUBO) objective:
\vspace{-0.5em} 
\begin{equation}
\min_{\mathbf{x}} H(\mathbf{x}) = \sum_{i=1}^n h_i x_i + \sum_{i < j}^n J_{ij} x_i x_j, \quad x_i \in \{0, 1\},
\label{eq:qubo_selection}
\end{equation}
\noindent where $\mathbf{x} = [x_1, \dots, x_n]^\top$ is a vector of binary decision variables, with $x_i = 1$ indicating that the $i$-th sample is selected for the current batch. The linear coefficients $h_i$ quantify individual uncertainty (e.g., via confidence or regression variance), while the quadratic couplings $J_{ij}$ represent pairwise similarities to enforce batch diversity.

\subsection{Samplers}
We provide two samplers based on the Kaiwu API. Both optimizers receive an Ising interaction matrix $J$ and an external field vector $h$, which are converted from the current weights and biases of the BM. These parameters define the energy landscape, and the optimizers perform sampling by searching for solutions that correspond to the global minimum. Repeated invocations generate sufficient samples to estimate negative phase statistics, which are then used to calculate gradients for updating the neural network parameters.

\paragraph{SA Optimizer / Sampler.} As a classical benchmark, we implemented a CPU-based simulated annealing optimizer. It searches for the global minimum within complex energy landscapes by simulating a thermodynamic cooling process. This mechanism allows the system to escape local optima, eventually generating samples that approximately follow the Boltzmann distribution defined by the model. Running entirely locally without specialized hardware, it is primarily used for rapid debugging, hyperparameter optimization, and performance benchmarking against physical solvers.

\paragraph{CIM Optimizer / Sampler.} This optimizer requires physical hardware access for quantum acceleration. The Ising matrix is transmitted via API to a cloud-based CIM. It completes the sampling process by reading the underlying optical quantum physical evolution as a set of low-energy spin configurations, which are then returned as Boltzmann samples. In our implementation, we de-duplicate the results of 2,000 optimization runs and apply Top-$K$ filtering to mitigate noise and obtain high-quality Boltzmann sampling results.

\subsection{Model Zoo and Loss}
We present the implementation of energy-based generative models within the KPP framework across various tasks.

\paragraph{QBM.} To standardize energy model implementations, we developed \texttt{AbstractBoltzmannMachine}, an abstract base class inheriting from \texttt{torch.nn.Module}. This class encapsulates fundamental BM operations and decouples energy definitions from sampling mechanisms, facilitating seamless extension. It defines a canonical objective function to calculate the average energy difference between observed data (positive phase) and the model distribution (negative phase), where the gradient serves as an approximation of the negative log-likelihood. Furthermore, it provides an extensible interface to map neural network parameters into an Ising interaction matrix, enabling integration with diverse sampling backends, including the CIM. Based on this abstraction, we have implemented several models, including classical BM, RBM, DBN, QBM-VAE, QGAN, and Q-Diffusion.

\paragraph{QBM-VAE. } QBM-VAE \shortcite{wang2025quantumboostedhighfidelitydeeplearning} is a hybrid quantum-classical deep learning architecture designed to introduce a Boltzmann distribution as the latent space prior. This approach captures physical correlations far more complex than those accessible via standard Gaussian assumptions. 
To interface with the binary states of quantum hardware, the latent space is constructed in a discrete form. Since direct discrete sampling is non-differentiable and prevents gradient propagation, the implementation utilizes an encoder $q_{\varphi}$ (parameterized by weights $\varphi$) to first map the high-dimensional input $x$ to $n$-dimensional Bernoulli probabilities $q_{l}$, representing the probability of each bit being 1.
Subsequently, a random noise $\rho_{l}$ drawn from a uniform distribution $U(0, 1)$ is introduced. Combined with a scale factor $\beta$ (fixed at 0.5), it is processed through the following reparameterization sampling formula to generate the continuous relaxation variable $\zeta_{l}$:
\vspace{-0.5em}
\begin{equation}
\zeta_{l}(\rho_{l},q_{l}) = \frac{1}{\beta} \log \left[ \frac{\max(\rho_{l} + q_{l} - 1, 0)}{q_{l}}(e^{\beta} - 1) + 1 \right],
\end{equation}

This $\zeta_{l}$ serves as a ``smooth proxy'' for the binary latent variable $z \in \{0, 1\}$. By approximating discrete states numerically while maintaining differentiability, it ensures that gradients can flow back to update the encoder via backpropagation. The physical stability of the binary latent variable $z$ is governed by the energy function $E_{\theta}(z)$, which forms the foundation of the Boltzmann prior $p_{\theta}(z) \equiv e^{-E_{\theta}(z)}/Z_{\theta}$.
The model is trained by minimizing the KL divergence to align the posterior distribution with the Boltzmann prior:
\vspace{-0.5em}
\begin{equation}
D_{\text{KL}}(q_{\varphi}(z|x) \parallel p_{\theta}(z)) = \mathbb{E}_{q_{\varphi}}[\log q_{\varphi}] + \mathbb{E}_{q_{\varphi}}[E_{\theta}(z)] + \log Z_{\theta},
\end{equation}

The log-partition function term $\log Z_{\theta}$ is computationally intractable in classical computing due to its exponential complexity ($2^{n}$ states). Therefore, this term is offloaded to CIM for efficient solution via physical sampling.
\vspace{-0.1em} 
\paragraph{Q-Diffusion.} This represents a novel integration of energy models in generative tasks. Specifically, the QBM is embedded into the loss function of Diffusion-based Large Language Models, revealing a new paradigm for leveraging energy-based priors in continuous generative processes. Existing discrete diffusion models suffer from a fundamental limitation: at each denoising step, they typically factorize the joint distribution $p_{\theta}(x_0|x_t)$ into a product of independent token-wise distributions, thereby neglecting contextual correlations among tokens within the sequence. To address this issue and enable controllable generation within high-dimensional discrete semantic spaces, we propose a residual energy diffusion framework driven by Quantum Boltzmann Machines (QBMs). Under this framework, the generation process no longer relies solely on the pretrained diffusion model; instead, it is guided by a joint distribution determined collaboratively by the diffusion model and the QBM energy function, revealing a novel integration paradigm for EBMs in generative tasks \shortcite{xu2024energy}. The denoising distribution is modeled in the following residual energy formulation:
\vspace{-0.7em}
\begin{equation}
    p_{\theta,\phi}(x_{0}|x_{t}) = \mu_{\theta}(x_{0}|x_{t}) \frac{\exp(-E_{\text{QBM}}(x_{0}))}{Z_{\phi}(x_{t})}
\end{equation}

where $\mu_{\theta}$ denotes the pretrained diffusion proposal distribution, and $Z_{\phi}$ represents the normalization constant (partition function). To leverage the advantages of quantum computing and quantify semantic constraints, we substantiate the energy function $E_{\text{QBM}}$ in the form of an Ising model, mapping text sequences to binary spin states to facilitate efficient energy evaluation and sampling on quantum hardware.
Addressing the intractability of calculating the exact partition function $Z_{\phi}$, we adopt targeted circumvention strategies during the training and generation phases, respectively.
During generation, we utilize a local approximation strategy. In the importance sampling step, we compute the sum of weights for only $k$ candidate samples generated by the diffusion model, using this ``local sum'' to substitute the global $Z$ for normalization. 
During training, we employ the Noise Contrastive Estimation (NCE) \shortcite{gutmann2010noise} principle to design the loss function $\mathcal{L}$. The ingenuity of NCE lies in transforming density estimation into a binary classification problem, eliminating the need for explicit calculation of $Z$ during parameter updates. Specifically, we optimize the QBM parameters by discriminating between real samples $x^+$ and negative samples $x^-$ generated by the diffusion-based Large language models:
\vspace{-0.5em}
\begin{equation}
\begin{split}
    \mathcal{L} &= \mathbb{E}_{\substack{x_0 \sim p_{data}, \\ x_t \sim p(x_t|x_0)}} {\biggl[} \mathbb{E}_{x^+ \sim q(x|x_0)} \Bigl[ \log \frac{1}{1+\exp(E_{\text{QBM}}(x^+))} \Bigr] \\
    &\quad + \mathbb{E}_{x^- \sim q_\theta(x|x_t)} \Bigl[ \log \frac{1}{1+\exp(-E_{\text{QBM}}(x^-))} \Bigr]{\biggr]} .
\end{split}
\end{equation}
\vspace{-1.7em}
\section{Results}
Our model achieved SOTA performance across the vast majority of metrics—including batch effect Correction, representation clustering, and classification—on all single-cell representation datasets (complete results omitted due to space constraints). 
\vspace{-0.6em}
\begin{table}[htbp]
\centering
\vspace{-0.6em}
\renewcommand{\arraystretch}{0.9}
\begin{tabular}{l|cc|cc|cc}
\toprule
 & \multicolumn{2}{c|}{\textbf{Immune}} & \multicolumn{2}{c|}{\textbf{HLCA}} & \multicolumn{2}{c}{\textbf{Lung}} \\
\textbf{Model} & Bio. & Batch & Bio. & Batch & Bio. & Batch \\ \midrule
VAE \shortcite{kingma2013auto} & 0.55 & 0.36 & 0.54 & 0.34 & 0.56 & 0.39 \\
scVI \shortcite{ergen2025scvi}        & 0.65 & 0.59 & 0.72 & 0.44 & 0.52 & 0.47 \\
LDVAE  \shortcite{ergen2025scvi}      & 0.55 & 0.58 & 0.65 & 0.43 & 0.53 & 0.47 \\
scPoli \shortcite{de2023population}     & 0.65 & 0.59 & 0.71 & 0.43 & 0.55 & 0.47 \\ \midrule
\textbf{QBM-VAE} & \textbf{0.69} & \textbf{0.61} & \textbf{0.76} & \textbf{0.48} & \textbf{0.67} & \textbf{0.49} \\ \bottomrule
\end{tabular}
\vspace{-0.6em}
\caption{Performance comparison on selected datasets}
\end{table}
\vspace{-0.6em}

On the OpenWebText dataset, our approach outperformed established models such as Masked Language Diffusion Model (MLDM) \shortcite{he2023diffusionbert} and Score Entropy Discrete Diffusion (SEDD) \shortcite{lou2023discrete}. This evidence confirms QBM's pivotal role in reducing the perplexity of Diffusion-based LLMs.

\begin{table}[h]
    \vspace{-0.6em}
    \centering
    \begin{tabular}{c|ccc}
        \hline
        \textbf{Model} & \textbf{Ours} & SEDD & MLDM  \\ \hline
        \textbf{PPL} & \textbf{22.54} & 24.56 & 23.83 \\ \hline
    \end{tabular}
    \vspace{-0.6em}
    \caption{OpenWebText results for the first three models}
    \label{tab:openwebtext_results}
\end{table}
 \vspace{-1em}
\newpage
\bibliographystyle{named}
\bibliography{ijcai26}

\end{document}